# Timescales of Massive Human Entrainment

R. Fusaroli[1,2*], M. Perlman[3], A. Mislove[4], A. Paxton[5], T. Matlock[5], R. Dale[5]

Author affiliations:

[1] Interacting Minds Center, Aarhus University, Denmark

[2] Center for Semiotics, Aarhus University, Denmark

[3] Department of Psychology, University of Wisconsin, Madison

[4] College of Computer and Information Science, Northeastern University, Boston, MA

[5] Cognitive and Information Sciences, University of California, Merced

* Corresponding author:

    Dr. Riccardo Fusaroli
    Building 1485, Jens Chr Skous Vej 2, 8000 Aarhus.
    E-mail: Fusaroli@gmail.com

# Abstract

The past two decades have seen an upsurge of interest in the collective behaviors of complex systems composed of many agents entrained to each other and to external events. In this paper, we extend concepts of entrainment to the dynamics of human collective attention. We conducted a detailed investigation of the unfolding of human entrainment—as expressed by the content and patterns of hundreds of thousands of messages on Twitter—during the 2012 US presidential debates. By time locking these data sources, we quantify the impact of the unfolding debate on human attention. We show that collective social behavior covaries second-by-second to the interactional dynamics of the debates: A candidate speaking induces rapid increases in mentions of his name on social media and decreases in mentions of the other candidate. Moreover, interruptions by an interlocutor increase the attention received. We also highlight a distinct time scale for the impact of salient moments in the debate: Mentions in social media start within 5-10 seconds after the moment; peak at approximately one minute; and slowly decay in a consistent fashion across well-known events during the debates. Finally, we show that public attention after an initial burst slowly decays through the course of the debates. Thus we demonstrate that large-scale human entrainment may hold across a number of distinct scales, in an exquisitely time-locked fashion. The methods and results pave the way for careful study of the dynamics and mechanisms of large-scale human entrainment.

# Introduction

Interest in the collective behaviors of complex systems composed of many agents has dramatically increased over the past couple of decades. This interest may stem in no small part from a new ability to measure and model collective behaviors. In a canonical case, Strogatz and Stewart [1] highlight firefly behavior as illustrative of fundamental principles underlying entrained systems [2,3]. In parts of Southeast Asia, one may happen upon a sea of fireflies, in which each firefly's intrinsic oscillatory dynamics have become entrained to others around it. The result is a large-scale collective behavior: The fireflies fire in sync in an impressive display brought on by subtle mutual influences. They are *entrained* in that they match their behavior to the temporal structure of events in the environment [4-6]. This process might involve elements of reciprocal influence between individual agents as in the case of the fireflies, or it might depend predominantly on external environmental events. The firefly model has inspired the investigation of entrainment across many physiological and technological phenomena, from neuronal firing to electric power networks [7]. However, it is still unclear how complex cognitive agents, such as human beings, might also exhibit patterns of large-scale entrainment.

In this paper we employ a series of massively shared media events to examine the entrainment of human collective attentional behavior at several time scales. We analyzed the three 2012 US presidential debates between Barack Obama and Mitt Romney, altogether watched by 192 million viewers — and the associated use of Twitter, a popular social media service. These events were thus (a) *shared at a massive scale*, and, via Twitter, (b) *induced the rapid spread of social behavior across a network of agents*. We time locked the corresponding Twitter data with video of each debate to match precise behaviors in the debates with the second-by-second rate of tweets involving mentions of the candidates. With these two time series in hand, we examined whether human behavior is entrained at three different time scales: i) short-term entrainment to conversational dynamics; ii) slower entrainment to salient contents of the debates; and iii) long-term entrainment to the duration of the debates. We define statistical models that can capture the aggregate tendencies of human behavior at these different scales, and test these on each debate to assess whether the effects generalize across events. The findings show massive behavioral entrainment in humans, which is intrinsically multi-scale and reproduces across events (the three debates).

## A massively shared event: US presidential debates

There are good reasons to choose the US presidential debates as our arena for exploring large-scale human entrainment. Since the televised debates of Kennedy and Nixon in 1960, they have attracted the attention of a hundred million or more television viewers each election cycle. The enormous magnitude of public attention has turned the debates into major events in the US Presidential elections, as candidates have the chance to sway millions of voters through the discussion of controversial issues and planned policies [9-11]. In addition to their massive television viewership, the most recent 2012 US Presidential debates—between candidates Barack Obama and Mitt Romney—were notable in the extent to which viewers were not just passive spectators isolated in front of a television set. Through the use of social media like Twitter and Facebook, millions of viewers participated in a global dialogue in which they generated tens of millions of interactive messages in real-time response to the debates.

The presidential debates present many salient aspects to public attention. Commentary on the debates emphasizes the highly competitive conversational interactions, dense with retorts, reciprocal interruptions and struggles for keeping or taking the floor [12-15], with much space devoted to assessing which candidate acted most presidentially [16-21]. Other studies have emphasized the content of the debates and how candidates frame the issues that are discussed [11,22,23], not least indicating the role of debates in creating widespread memes [24]. Finally, the debates, as any other large event, have a natural development as they warm up, reach their peak and then fade as they lose their novelty [25].

## A massively social behavior: the Twitter "gardenhose" stream

The recent development of massive social media networks yields a prime forum in which to examine the phenomenon of human collective entrainment. The use of social media technologies enables people to extend the existing constraints on the distance, timing, and connectivity of communication, facilitating the rapid cascade of information across the digital networks [8]. To investigate the impact of the presidential debates on human behavioral entrainment, we employed Twitter, a popular micro-blogging platform that launched in 2006. Twitter is widely used by marketers, public authorities, and the general public and has become a major mechanism for the rapid spread of information. As such it offers an unprecedented window into how large populations collectively experience and respond to a wide range of real-world events [26]. Researchers have used social media to describe—and sometimes anticipate—epidemics, earthquakes, stock options, the effect of time and weather on mood, reality show outcomes, and political elections [27-36]. Little is known, however, about the precise temporal dynamics through which the use of online social media reflects and interacts with the unfolding action of massively shared events. We chose to investigate these dynamics with Twitter because of the near-instantaneous nature of its message: Its short format (140 characters per message) and widespread integration with mobile devices facilitates fast messaging and reactions. Twitter provides a grasp of the precise temporal dynamics of how real-world events drive and resonate with human social behavior.

## The dynamics of human collective entrainment: Three time scales

The purpose of this study was to explore human entrainment to the Presidential Debates through Twitter. Human social entrainment is arguably more complex than that of other species; events that reflect the sophisticated format of human interaction may shape entrainment in distinct ways. We thus hypothesized that the fine-grained conversational dynamics of the debates would directly drive and constrain Twitter discourse concerning the events at (at least) three time-scales of interest.

i) *Interactional entrainment*: We hypothesized that assertive behaviors—keeping the ground, interrupting the adversary, and so on —would strongly impact Twitter mentions and lead to higher rates of tweeting about the respective candidate. Thus candidates would generate tweets as they interrupted their opponent and asserted their turn, and they would continue to generate tweets for as long as they maintained the floor. This hypothesis was motivated by political and media studies suggesting that presidential debates are employed as heuristic or judgmental shortcut for viewers to assess future presidential performance [16,17]. Both experimental settings and real life analyses showed that human beings tend to perceive and support leadership in individuals with extroverted personalities [37,38] and relatedly in those who display assertiveness, boldness, initiative, proactivity, and risk-taking [39-42]. Corroborating this view is extensive coverage by the news media of the interactional style of the candidates—who behaved more presidentially, who was being defensive—with victory often defined in terms of the level of interruptions and direct confrontation [20,21][1].

ii) *Content entrainment*: Besides this ebb-and-flow dynamics of interaction, debates are also rife with pointed or "salient" remarks that propagate through social media—often as "memes" that cascade through communications in forums like Twitter [44]. Indeed, viewers pay attention to the contents of the debates, focusing their attention on particularly salient, amusing or controversial elements [24]. We hypothesized that viewers would react to these salient events, however, in different ways than to the candidates' conversational dynamics. Content entrainment is likely to require more intensive cognitive processing and therefore happen at longer time scales. Moreover, interest in salient events is expected to be partially self-sustaining: Once a high level of attention has been raised, the tweets produced will help maintain the attention on the topic, although the debate might have moved on.

---

[1] For the current paper, we did not consider emotional valence of attention. Instead, we hypothesized that display of assertiveness would capture the attention of viewers, irrespective of whether that attention was positive or negative. We leave the evaluation of judgments, emotional valence and more sophisticated clustering in the response to assertiveness display to future studies. 43. Lin Y-R, Margolin D, Keegan B, Lazer D. Voices of victory: A computational focus group framework for tracking opinion shift in real time; 2013. International World Wide Web Conferences Steering Committee. pp. 737-748.

iii) *Long-term attention decay*: Finally, despite the relatively longer scale of content entrainment, attention and interest are unlikely to be sustained for a long period, being subject to bursts and decays [8]. Therefore, we expected the general interest in the debate to decay after an initial burst, thus showing long-term attentional dynamics.

Below we demonstrate how the entrainment of Twitter behavior to the Presidential Debates is aptly characterized by these three time scales, both individually and in a multi-scale model.

# Materials and Methods

## Analysis of the debate

There were three 2012 US presidential debates between former Massachusetts Governor Mitt Romney and incumbent US President Barack Obama. The first took place on October 3rd at University of Denver, Denver, Colorado; the second on October 16th at Hofstra University, Hempstead, New York; and the third on October 22 at Lynn University, Boca Raton, Florida. Each debate lasted about 90 minutes.

The audio recordings and transcripts of the three debates between President Barack Obama and Governor Mitt Romney were collected from National Public Radio ([www.npr.org](www.npr.org)). The transcripts were cleaned and edited to better reflect the audio files. Through careful listening supplemented by an in-depth examination of the waveform and automated analysis of variations in pitch and intensity using Praat [45] and MATLAB (Mathworks Inc.), we individuated start and end time at a 10-millisecond scale for each speech turn as well as interruptions and a selection of salient moments discussed in popular media after the debate events (see Figure 1). This was performed blind from any inspection of the Twitter data (see below). By identifying the precise timestamp of the debate onset, we time-aligned the Twitter data and the debate data (see Figure 2).

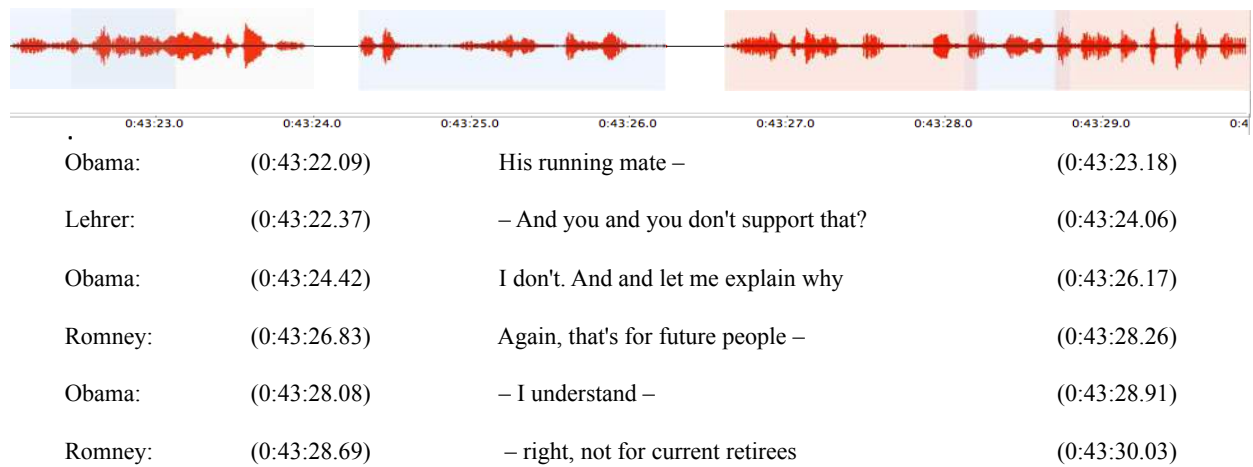

| | | | |
|---|---|---|---|
| Obama: | (0:43:22.09) | His running mate – | (0:43:23.18) |
| Lehrer: | (0:43:22.37) | – And you and you don't support that? | (0:43:24.06) |
| Obama: | (0:43:24.42) | I don't. And and let me explain why | (0:43:26.17) |
| Romney: | (0:43:26.83) | Again, that's for future people – | (0:43:28.26) |
| Obama: | (0:43:28.08) | – I understand – | (0:43:28.91) |
| Romney: | (0:43:28.69) | – right, not for current retirees | (0:43:30.03) |

**Figure 1. Excerpt of the waveform and related transcript from the first presidential debate.** Blue highlighting indicates Obama speech turns, red Romney's and grey Lehrer's (the moderator). The transcripts were retrieved from the National Public Radio website, cleaned and edited to better reflect the audio files. Start and end time of each speech turn as well as interruptions are aligned to the debate combined careful coding and automated processing.

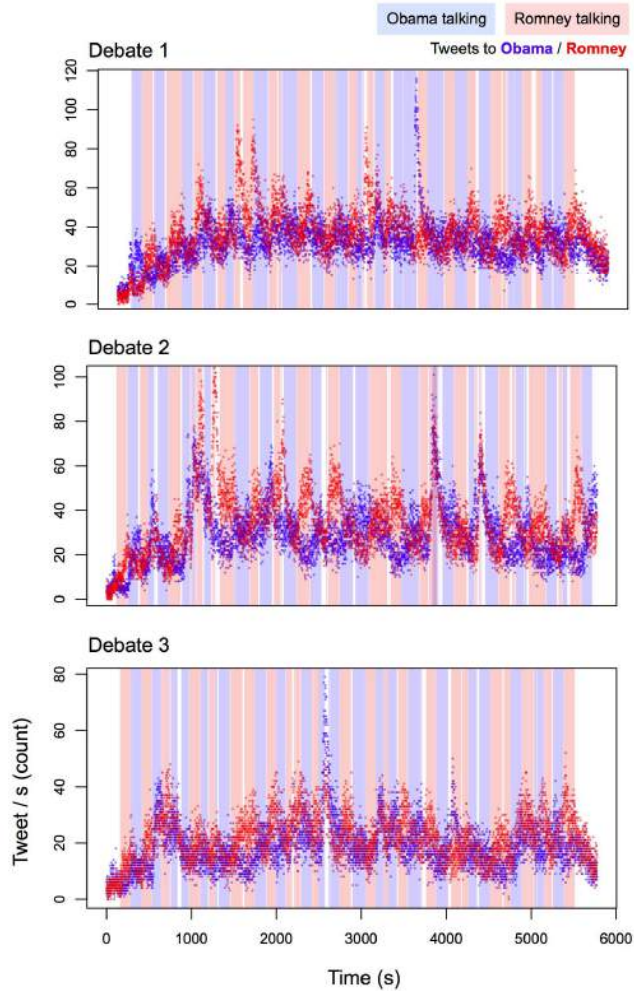

**Figure 2. Tweet rate and turn-taking during the presidential debates.** Light red and blue rectangles are periods of time during which candidates were speaking during the debates. Darker red and blue dots represent per-second tweet rate mentioning the corresponding candidates. Visual inspect reveals relatively periodic patterns of Twitter mentions that seem to be cued by turn onset. Plots include both tweets and retweets in the tweet / s rate.

## Analysis of the tweets

The Twitter data consisted of a random sample of approximately 10% of all public tweets ( "gardenhose" stream), collected during each 90-minute presidential debate. The Twitter data collected as part of this study currently resides on and is archived by co-author Mislove's research cluster at Northeastern University. While the data source (Twitter's streaming service) is publicly available, Twitter's Terms of Service prevent making the raw tweets available. Instead, we make the list of unique

6 of 21

tweet identifiers (tweetIDs) publicly available (on http://www.ccs.neu.edu/home/amislove/obama-romney/), similar to previous studies of Twitter and Twitter-based benchmarks.

We filtered tweets to select only those that mentioned "Obama" or "Romney," either in the text or in their hashtag, and we excluded those containing URLs (to exclude spambot-generated tweets). This resulted in 713,642, 686,805, and 406,242 tweets for the first, second, and third debates, respectively. Each set of tweets was generated by a large number of unique user accounts: 442,368, 413,537, and 255,644 accounts respectively for each debate (see Table 1). "Retweets" (i.e., when another Twitter user merely reposted the original message) were omitted from the analysis, which ensured these patterns were not simply generated by repetitions of the same messages [46]. However, analyses including retweets show similar robust patterns (see Supplementary Figures S1 to S4).

**Table 1. Basic descriptive statistics of Twitter data collected for the debates**. Sum of "Obama" and "Romney" may exceed total tweet count because tweets can mention both of them.

| Debate | Total tweets | Retweets | Mean tweets / sec (SD) | "Obama" | "Romney" |
|---|---|---|---|---|---|
| 1 | 713642 | 381797 | 110.4 (47.2) | 411391 | 468583 |
| 2 | 686805 | 368010 | 104.5 (47.9) | 375506 | 462159 |
| 3 | 406368 | 212262 | 63.0 (27.8) | 231778 | 266801 |

## Statistical analysis of combined debate and Twitter data

We assessed the impact of debate events on human entrainment as measured in tweet rate per second at three key time scales. An overlay of tweet rate per second and turn-taking for each debate is shown in Figure 2. We first modeled each scale individually. Then we built a multiple regression model including all three time scales to assess their relative and overall predictive power for public attention. We hypothesized the three debates to display the same trends: statistically significant attentional entrainment at the three time scales. To ensure effects were not driven by one debate only, we fitted each model to each single debate and report them separately. To further ensure the generality of our results after fitting the full multiple regression model to the first debate, we employed it to predict attentional entrainment in the other two debates. Full details of the analyses are reported in the following paragraphs. All models were developed with the lme4 and MuMIn libraries in R, and the R code is available in the Supporting Information File S1.

**Interaction: Turn-taking and interruptions**

The first time scale was modeled on the turn-taking dynamics, using number of tweets per second (measured at a 1-second scale) as the dependent variable and "speaker", "speaking time", and "interruption" as independent variables. Speaker was a dichotomous factor indicating which speaker is holding the floor. Speaking time was a measure of how long the speaker had been speaking in the current speech turn. Interruption was a dichotomous factor indicating whether the current speaker had interrupted his interlocutors to gain the floor. Linear mixed effects models were used to test these patterns for each debate. The first model included a main effect for speaker (candidate vs. others), duration of speaking in each speech turn, and an interaction between these two fixed factors. The models included a random effect for turn number, along with nested slopes for both candidate identity and time within turn number. The second model built on the first model by including interruption as an additional fixed factor.

**Content: Momentary salient events**

To investigate the second time scale, the impact of content, we chose three distinct salient events that took place in the interaction. These events, which quickly evolved into Internet "memes," were identified based on popularly discussed comments by the candidates. We chose one salient moment per debate: Romney declaring "I love Big Bird" in the first debate, Romney mentioning that he received "whole binders full of women" in the second debate, and Obama noting that the army had fewer "bayonets" in the third debate. Each of these events spread rapidly on the Internet, becoming the dominant topics of debate-related Twitter conversations and online searches for each of them totaled hundreds of thousands of mentions [24].

We expected attention to salient events to have partially self-sustaining dynamics. When enough tweets are produced on a given topic, they should keep public attention focused on that topic, although the debate might have moved on. To estimate how long a salient event can be expected to influence overall tweet counts, collective attentional entrainment at this scale was modeled as an exponential decay function coupled to a sigmoid. This serves as a simple mathematical model for a meme. The decay component relates to the fall from a burst of mentions due to novelty or salience of the event, $N(t) = e^{-\lambda t}$, with $\lambda$ reflecting the decay rate. If that saliency achieves a particular prominence, or threshold, then the continuing attention to the event may sustain it as a meme, which could be characterized as a rapid-onset sigmoid function, $M(t) = 1 / (1+e^{-m(t-s)})$, where $s$ is the point (in seconds) at which tweet rate is increasing maximally for the "meme," and $m$ reflects the slope of that rate. The following product of these two functions captures the general patterns seen in Figure 5:

(Eq. 1) $M(t) [N(t)-b]$

$b$ is the mean base tweet rate observed in the final 100s of the data, reflecting the stable sustained tweet rate after the initial rapid decay. The model was fit to the three events by performing a simple parameter search within reasonable ranges of $\lambda$, $s$, and $m$, and choosing parameter values that maximized the correlation between the model and the observed data.

**Long-term attention: The whole debate**

The longest timescale was represented as a quadratic time term that rises from the onset of the debate, and drops at its end. This is motivated by the notion that human social responsiveness to the debate will itself be driven by the onset and offset of the massively shared event. We tested for the impact of long-term attention by employing a linear multiple regression model with tweets per second (measured at a 1-second scale) as dependent variable and a second-order polynomial as independent variable to account for linear and quadratic temporal development. The presence of decay in the second half of the debate was further tested by assessing the fit of the quadratic term alone (which involves only predicted decay in the second half of the inverse quadratic function).

**Multi-scale dynamics: Predicting public attention**

We combined the three time scales variables into one regression model per debate that predicted overall rate of tweets. Thus, we employed number of tweets per second as the dependent variable, and "speaker duration", "interruption, " "salient moment" and "quadratic time" as independent variables. As shown below, each time scale contributes uniquely to the model, suggesting that entrainment of large-scale social attention is complex and driven by several time-varying factors. Finally, we tested whether the model generated from the first debate would generalize to predict tweet rates in the second and third debates. We chose not to include salient events in this last test as their analysis relied on post-hoc assessment of which events went viral and therefore would not be easily generalizable to new debates.

# Results

**Interactional Entrainment 1: Tweet mentions co-vary with speaker**

Twitter activity was tightly time-locked with turn-taking exchanges in each of the debates (Figure 3). When one candidate started to speak, tweet rate increased for that candidate within seconds of the turn switch. The models for debates 1 to 3 explained at least 10% of the variance, with the tweet rate of debate 2 being the best explained by the model, at over 30% of its variance, for both Obama- and Romney-centered attention (all marginal $R^2$'s > .10). The models revealed both an increase in proportional mention for a candidate as he spoke, $ß$'s > .45, $t$'s > 3.3, $p$'s < .0001, and a significant interaction between speaker and speaking time, $ß$'s > .40, $t$'s > 4, $p$'s < .0001 (see Table 2). Thus the more a candidate spoke in each single speech turn, the more attention he received. This suggests that entrainment to the turn-taking structure of the debate is rapid, requiring only a few seconds to exert an observable influence on massive social attention. All three debates display the same significant factors, with analogous effect direction and size.

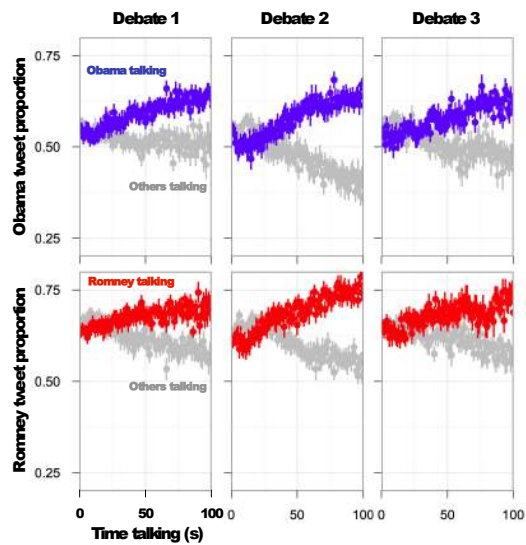

**Figure 3. Effects of taking and holding the ground on Twitter mentions.** Starting from the onset of each turn per candidate, plots show relative proportion of Twitter mention rises during that candidate's turn. While others are speaking, proportion mentions drops. Proportions are based on, for example, dividing mention to "Obama" divided by the sum of mentions to "Obama" and "Romney" together. Importantly, these plots only include original tweets, showing the anticipated effect is independent of retweets.

**Table 2. Speaking by candidates strongly invokes Twitter attention.** Estimates of $ß$ were calculated by standardizing all continuous variables. Across all three debates, speaker mention substantially drives attention (tweet mention). $ß$ and $t$'s are reported as Obama / Romney, as a model was devised to test the effect of each speaker's turns.

| Debate | *Speaker $ß, t$* | *Speaking time $ß, t$* | *Speaker x Speaking $ß, t$* |
|---|---|---|---|
| 1 | Obama: 0.75, 4.6 | Obama: -0.30, -4.3 | Obama. 0.70, 6.3 |
|   | Romney: 0.67. 4.8 | Romney: -0.37, -5.1 | Romney: 0.69, 7.4 |
| 2 | Obama: 1.2, 8.6 | Obama: -0.50, -6.8 | Obama: 0.98, 9.1 |
|   | Romney: 1.0, 8.2 | Romney: -0.42, -5.4 | Romney: 0.98, 9.1 |
| 3 | Obama: 0.47, 3.4<br>Romney: 0.60, 4.5 | Obama: -0.21, -2.7<br>Romney: -0.22, -3.1 | Obama: 0.50, 4.3<br>Romney: 0.43, 4.0 |

**Interactional Entrainment 2: Tweet rate increases with conversational interruptions**

Tweet rate was also influenced by interruptions, which significantly increased Twitter mentions of both candidates. Figure 4 shows the tweet rate for both candidates and moderator together when their turns were interruptions or not. Numerous interruptions took place in the debates and were of varying lengths (Table 3).

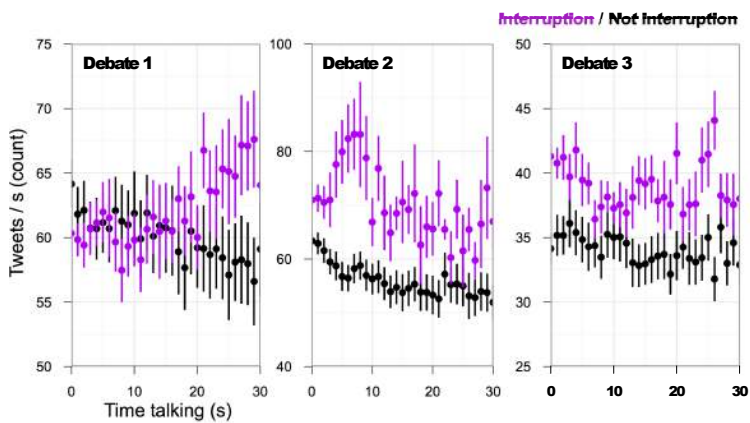

**Figure 4. Effects of interruptions on Twitter mentions.** At the onset of speaking, results show that the volume of tweets increases when that spoken turn is in the form of an interruption. Each panel represents the results from one of the debates. Importantly this figure only shows original tweets, omitting retweets.

**Table 3. The number of interruptions identified in each debate.** Duration range is the minimum / maximum length of turn identified as an interruption. The final three columns identify interruption counts within speaker.

| Debate | Turn count | Interruptions | Duration range (s) | Obama | Romney | Moderator |
|---|---|---|---|---|---|---|
| 1 | 214 | 115 | 0.1 – 130.6 | 23 | 45 | 47 |
| 2 | 266 | 105 | 0 – 208.7 | 39 | 37 | 29 |
| 3 | 190 | 117 | 0.1 – 117.7 | 41 | 45 | 31 |

Results revealed a general increase in the mention of both candidates during interrupting events. Using a mixed effects model similar to the prior analysis, all debates show a reliable contribution of interruption, with marginal $R^2$'s = .07, .02, and .12, for debates 1-3, respectively. Though the effect of interruptions is much smaller, all three debates show a significant coefficient for the interruption term, $\beta$'s > .50, $t$'s > 1.9, $p$'s < .05 (see Table 4). All three debates display the same significant patterns, with analogous effect direction and size.

**Table 4. Interruption by candidates increases Twitter activity.** Estimates of $\beta$ were calculated by standardizing all continuous variables. Across all three debates, interruptions drive debate attention (overall tweet rate).

| Debate | *Interruption $\beta$, t* | *Speaking time $\beta$, t* | *Interruption x Speaking time $\beta$, t* |
|---|---|---|---|

| 1 | 0.72, 2.0 | -0.20, -3.8 | 0.80, 2.5 |
| 2 | 0.55, 2.5 | -0.07, -0.64 | 0.11, 0.72 |
| 3 | 0.94, 4.2 | -0.08, -1.34 | 0.29, 1.6 |

**Content entrainment: Twitter bursts to "memes"**

Twitter behavior was influenced by the occurrence of salient momentary events that took place during the debates. Focusing on tweets containing the root terms "big bird" (10,076 mentions), "binder" (2,889), or "bayonet" (5,458), we analyzed the temporal development of Twitter behavior following the precise onset of each event. Our analysis shows that Twitter behavior displayed a remarkably similar temporal profile for each of these events. The first mention of the terms occurred within 11 seconds, and tweet rates peaked at about one minute after its onset, followed by a slow decay over the next few minutes (Figure 5). Using the model of meme initiation and propagation we described in the previous section (Eq. 1), we model these temporal profiles in Figure 6. Distinct meme-like events can be modeled with the same functional form, and model parameters may serve to characterize subtle distinctions among them.

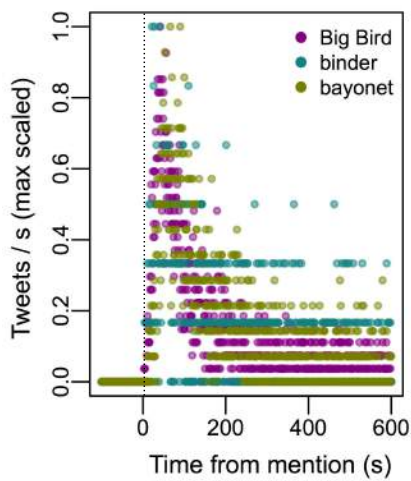

**Figure 5. The temporal profile of public attention to salient events.** At the onset of a salient event, mention of the word (in the context of either "Obama" or "Romney") rapidly rises within 10 seconds (left panel). Mentions are max scaled to facilitate comparison. Right panel shows retweets separately from original tweets, showing the expected delay. Interestingly, these salient events show distinct temporal signatures in their onset and rise to maximum, both in the profile of tweets and retweets. For original tweets, first mention for Big Bird, binder, and bayonet respectively is 4, 5, and 11 seconds; their maximum is achieved at 42, 23, and 67 seconds. In the retweet data, this is lagged, with first retweets at 31, 14, and 17 seconds; maximum achieved at 99, 80, and 78 seconds, respectively for Big Bird, binder, and bayonet.

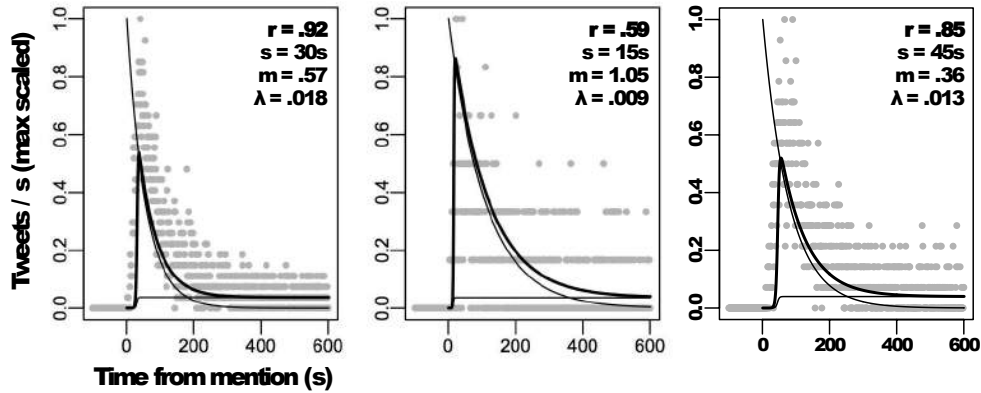

**Figure 6. A model of public attention to salient events.** The model of public attention reactions to salient events as fit to the three case studies: "Big bird", "binder" and "bayonet," from left to right. Note two interlocked timescales: a saliency/novelty followed by the establishment of a meme that sustains a base-level of continued attention.

**Long-term attentional decay**

We assessed the longer time scale of the debate itself, where we would expect both a gradual increase in attention, but one that trails off as the end of the debate approaches. Such a pattern is evident in Figure 2. To test this quantitatively, we used a second-order polynomial regression model, with first- and second-order time terms predicting overall tweet rate. Both are highly significant, and account for over 20% of the variance from the two terms alone, for each debate. The linearly increasing term is strongly significant, $ß$'s > .28, $t$'s > 20.0, $p$'s < .0001. However there appears to be a larger effect magnitude for the quadratic term, which specifies both a relative increase at the beginning of the debate and a *decrease* by the end of the debate, $ß$'s > .34, $t$'s > 25.0, $p$'s < .0001. This larger effect for the quadratic term holds for all three debates (see Table 5). Importantly, this was not driven just by the beginning of the debate, for which the nonlinear second-order term may be considered to fit better; the last half of the debate, which only includes the decay portion of the quadratic term, still shows a significant contribution of the decay term when included alone, $p$'s < .001 for all debates. All three debates display the same significant patterns, with analogous effect direction and size.

**Table 5. Long-term trends in the Twitter activity.** A OLS regression was used to predict tweet rate using a linear term representing the increasing time of the debate, and a quadratic term over the same time frame, which reflects and increase and then decrease. Both are highly significant, with the quadratic term in general having the larger effect size. That last column shows that the decay portion of the quadratic term still significantly predicts tweet rate when included alone. There is thus a longer timescale process of height activity then decay.

| Debate | $R^2$ | Linear term $ß$, $t$ | Quadratic term $ß$, $t$ |
|---|---|---|---|
| 1 | 0.29 | 0.34, 28.5 | -0.46, -38.1 |
| 2 | 0.23 | 0.36, 24.0 | -0.53, -35.2 |
| 3 | 0.16 | 0.28, 20.9 | -0.34, -26.0 |

**Regression model to test entrainment timescales**

The prior analyses demonstrated each time scale's relevance separately, and we wished to test in a simple way whether all of these factors contribute simultaneously to tweet rate. To do so, we developed a multiple regression model with all time scales as variables accounting for tweet rate. We factored in salient events, modeled as a decay function along with temporal variables for speaker, whether interruptions were taking place, and at the broadest scale, a quadratic term representing the start and end of the debate. In each debate, the full regression model accounted for almost 50% of the variance in tweet rate (see Figure 7). All variables also significantly ($p < 0.05$) and uniquely contributed to this variance (see Table 6). This regression analysis suggests that all time-varying properties that we have analyzed above contribute to the ebb and flow of public attention as reflected in tweet mentions. Put another way, the temporal variation in tweet rate may contain signatures of various time-scales of attentional processes taking place simultaneously in these massively shared experiences. These processes are influenced by broad exposure to the debate itself, by more local events, such as conversational interruptions and by the salient remarks that give rise to memes.

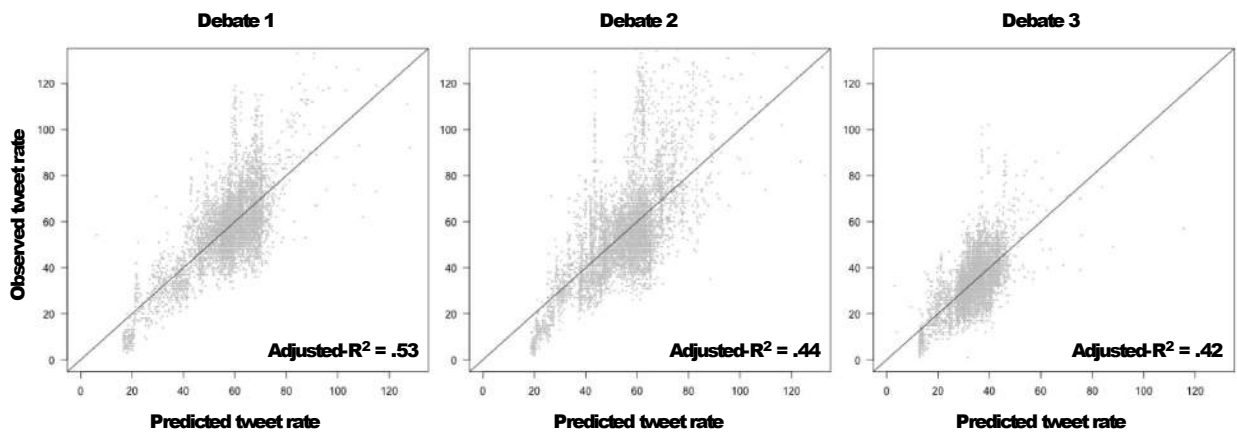

**Figure 7. Multiple regression fits for all 3 debates.** Variance accounted for by salient events, a quadratic time term, who is talking, and whether interruption is taking place accounts for between 42% and 53% of the variance in observed tweet rate. See text and Supporting Information for further details on the model.

**Table 6. Performance of simultaneous multiple regression models.** The variables developed in prior analyses were used in one multiple regression model, predicting tweet rate by a variety of factors. All contribute significantly. Proportion variance *uniquely* associated with each variable in the model is shown, by entering it last into the regression. Note that models include all interaction terms among our primary variables analyzed above (speaker, interruptions, etc.). See Supplementary Information for the full model specification.

| Debate | $R^2$ | Quadratic | Speaker duration | Salient | Interruptions |
|---|---|---|---|---|---|
| 1 | 0.53 | 0.20 | 0.06 | 0.11 | 0.02 |
| 2 | 0.44 | 0.12 | 0.04 | 0.11 | 0.04 |
| 3 | 0.42 | 0.09 | 0.03 | 0.14 | 0.06 |

Lastly, we used the model from the first debate to predict the tweet rate from the subsequent two debates. Can basic information about a debate—knowing the time point of the debate, whether one of the candidates is speaking, and whether one is interrupting the other—predict tweet rate from one debate to the next? Even with just these two timescales (speaker duration/interruption, the debate), the model from the first debate can capture about 10% or more of the variance in the second and third, $r$'s = .41, .32, respectively, $p$'s < 0001. A simple and efficient representation of basic conversational processes (speaker, interruption) and time terms (second-order polynomial) can significantly predict large-scale social attention across debates.

# Discussion

We hypothesized that the dynamics of a massively shared event—the 2012 US presidential debates—would be reflected in the second-by-second, larger-scale dynamics of public attention. Specifically, the generation of Twitter messages would exhibit entrainment to the debates at (at least) three time scales: short-term conversational dynamics, mid-term content and long-term attentional entrainment.

*i) Conversational entrainment:* Public attention and response are time-locked to the conversational dynamics (e.g. turn-taking, interruptions) of the debates. Within seconds of initiating their conversational turn, speakers generate increased Twitter mentions to themselves, with correspondingly fewer mentions to their opponent. Moreover, the longer the speaker holds the ground the greater the increase in attention he receives from the tweeting audience. Interrupting the adversary emphasizes this effect and increases attention on one's speech turn. In other words, collective attention is time-locked to cues of assertiveness and maybe even "presidentiality." It has to be restated that our findings are limited to how assertiveness display entrains viewers' attention and do not include more nuanced perspective on the emotional valence of the tweets, or even on possible clusters in response to different candidates.

*ii) Content entrainment:* In addition to a more immediate entrainment, we have shown slower dynamics as the public tunes its attention and elaborates on salient events. The first mention occurs with 11 seconds, mentions peak at 1 minute, and gradually fade over about 10 minutes time. The dynamics of this profile can be modeled as an interaction between the decrease of saliency of the event itself over time and the sustained interest generated by new mentions of the event on Twitter. This highlights the more demanding cognitive processing of actual semantic content, and the importance of intrinsic dynamics in the social media, which can keep a salient event alive beyond its instantiation in the debate. Interestingly, the results suggest that the salient event "binders," despite having a lower raw tweet rate relative to the other two salient events, had both the slowest decaying and more rapidly rising meme formation. This resonates with analyses by Lin et al. [24] showing that the "staying power" of a meme is not only related to the raw quantity of mention, but also other factors have to be taken into account, e.g. conversational vibrancy (i.e., the prominence of the tweeters involved) and the interactivity of their audience.

*iii) Long-term attention*: Not least, collective entrainment displays long-term dynamics with an initial increase as the debate unfolds, followed by a decrease as it nears its conclusion.

Altogether, our findings suggest that human collective entrainment is multi-scale. Each of these three scales contributes significantly and in non-overlapping ways to a multiple regression model predicting public attention in the form of Tweet rate.

While these results strongly indicate the presence of collective entrainment, they do not fully describe the complexity of human collective entrainment as many additional factors could and should be explored in future studies. Three dimensions in particular seem to be crucial for the current case study: i) emotional valence; ii) networks of political affiliation and pre-existing beliefs; and iii) impact on public opinion. Assertiveness and interruptions might generate positive appraisal as presidential qualities or they might be negatively assessed, and these reactions are likely to be mediated by political affiliation and pre-existing beliefs. Just as blogs cluster around political orientation [30], politically active Twitter users might primarily respond to their preferred candidate only, or modulate their appraisal of assertiveness and salient events so as to cast a good light on his behavior. Promising work has been done on automatically segmenting Twitter users according to their political orientation [43,47,48], on automated assessment of conceptual and emotional dimensions of political discourse and tweets [29,33,49,50], and on the impact of conversational dynamics between tweets[8,24]. Future work will also have to investigate the details of conversational entrainment through Twitter and its impact on public opinion.

We live in an age in which local events can be broadcast in real-time to hundreds of millions of people around the world, and in response, people can interact instantaneously with each other through the use of online social media. This qualitatively new capacity for communication is changing the nature of large-scale politics and the ability for people to coordinate action across the globe. The situation calls for the development of large-scale analysis and models that both characterize these emerging social dynamics as well as predict them. A growing number of studies are dedicated to identifying and categorizing events, such as earthquakes, and even successful and unsuccessful political speeches, according to the public attention dedicated to them [8,46,51-53]. Yet little is known about the dynamics of this local-global interaction. How does the unfolding action of debates and other broadcasted events impact real-time public attention and response in social media? By combining quantitative assessments of conversational dynamics [54-58] with the analysis of hundreds of thousands of Twitter messages, this study is the first to assess the unfolding impact of a single event on the large-scale dynamics of public attention. Our results highlight how the dynamics of a local conversation can entrain the communicative behavior of massive populations of spectators. They also demonstrate the value of fine-grained temporal analyses at different time scales in uncovering the powerful relationship between social media and public events.

# Conclusion

Collective and self-organizing behaviors are endemic to many social species, at many scales [59]. Entrainment is one frequently cited collective behavioral pattern, famous in fireflies [2], but found across numerous species, in murmurations of starlings, schooling in various fish species and more (see [60] for a review). Human communication might seem a smaller scale phenomenon, likely built on a foundation of dialogical and spatially limited interactive dynamics [61]. Recent studies, however, argue that large-scale entrainment dynamics could be observed, with local dialogical exchanges combining at a societal level and over time [62-64]. The advent of social media and information technologies allows humans to scale and speed up these dynamics to showcase massive and rapidly self-organizing patterns of entrainment. Indeed, our findings highlight that the massively-shared experience of a political event induces complex patterns of collective attentional entrainment: an exquisite time-locking of observed behavior with the structure of the political event itself, content entrainment with partially self-sustaining dynamics, and large-scale attention bursts and decays. Put simply, like "congregating fireflies", humans show massive sustained entrainment, across hundreds of thousands of individuals, in matters of seconds and minutes.

# Acknowledgments

The authors wish to thank Chris Kello, Chris Frith and Peer Christensen for useful suggestions, Skye Smith and Benjamin Riis for help in time-coding the transcripts.

# References

1. Strogatz SH, Stewart I (1993) Coupled oscillators and biological synchronization. Sci Am 269: 102-109.
2. Mirollo RE, Strogatz SH (1990) Synchronization of Pulse-Coupled Biological Oscillators. Siam Journal on Applied Mathematics 50: 1645-1662.
3. Pikovsky A, Rosenblum M, Kurths J (2001) Synchronization : a universal concept in nonlinear sciences. Cambridge: Cambridge University Press. xix, 411 p. p.
4. Ancona D, Chong C-L (1996) Entrainment: Pace, cycle, and rhythm in organizational behavior. Research in organizational behavior 18: 251-284.
5. Schmidt RC, Richardson MJ, Arsenault C, Galantucci B (2007) Visual tracking and entrainment to an environmental rhythm. Journal of Experimental Psychology Human Perception and Performance 33: 860-870.
6. Patel AD, Iversen JR, Bregman MR, Schulz I (2009) Experimental evidence for synchronization to a musical beat in a nonhuman animal. Current Biology 19: 827-830.
7. Dörfler F, Bullo F (2014) Synchronization in complex networks of phase oscillators: A survey. Automatica.
8. Lin YR, Keegan B, Margolin D, Lazer D (2014) Rising Tides or Rising Stars?: Dynamics of Shared Attention on Twitter during Media Events. PLoS One 9: e94093.
9. Jamieson KH, Birdsell DS (1988) Presidential debates: The challenge of creating an informed electorate. New York: Oxford University Press.
10. Carlin DP (1994) A rationale for a focus group study. In: Carlin DB, McKinney MS, editors. The 1992 presidential debates in focus. Westport, CT: Praeger. pp. 3–19.
11. Benoit WL, Hansen GJ, Verser RM (2003) A Meta-Analysis of the effects of viewing U.S. Presidential Debates. Communication Monographs 70: 335-350.
12. Grimshaw AD (1990) Conflict talk: Sociolinguistic investigations of arguments in conversations: Cambridge University Press.
13. Morello JT (1992) The "look" and language of clash: Visual structuring of argument in the 1988 Bush-Dukakis debates. Southern Communication Journal 57.
14. Benoit WL, McKinney MS, Lance Holber R (2001) Beyond Learning and Persona: Extending the Scope of Presidential Debate Effects. Communication Monographs 68: 259-273.
15. Pfau (2002) The subtle nature of presidential debate influence. Argumentation and Advocacy 38: 251-262.
16. Patterson ML, Churchill ME, Burger GK, Powell JL (1992) Verbal and nonverbal modality effects on impressions of political candidates. Analysis from the 1984 presidential debates. Communication Monographs 1992: 231-242.
17. Masters RD, Sullivan D (1993) Nonverbal behavior and leadership. In: Iyengar S, McGuire WJ, editors. Explorations in Political Psychology. Durham (NC): Duke University Press.
18. Sullivan DG, Masters RD (1994) Biopolitics, the media, and leadership: Nonverbal cues, emotions, and trait attributions in the evaluation of leaders. In: Somit A, Peterson SA, editors. Research in Biopolitics. Greenwich, CT: JAI Press. pp. 237–273.
19. Bucy EP, Grabe ME (2008) "Happy warriors" revisited: hedonic and agonic display repertoires of presidential candidates on the evening news. Politics Life Sci 27: 78-98.
20. Hamby P, Preston M, Steinhauser P (2012) 5 things we learned from the presidential debate. CNN.com.

21. Ward J (2012) Mitt Romney Versus Obama: 4 Key Moments From First Presidential Debate. The Huffington Post.
22. Clayman SE (1995) Defining moments, presidential debates, and the dynamics of quotability. Journal of Communication 45: 118-147.
23. Matlock T (2012) Framing Political Messages with Grammar and Metaphor-How something is said may be as important as what is said. American Scientist 100: 478.
24. Lin YL, Margolin D, Keegan B, Baronchelli A, Lazer D (2013) #Bigbirds never die: understaing social dynamics of emergent hashtags. Proceedings of 7th International AAAI Conference on Weblogs and Social Media: 370–379.
25. Schroeder A (2008) The Presidential Debates: Fifty Years of High-Risk TV: Columbia University Press.
26. Miller G (2012) Social Scientists wade into the Tweet stream. Science 333: 1814-1815.
27. Culotta A. Towards detecting influenza epidemics by analyzing Twitter messages.; 2010; New York. pp. 115-122.
28. Bollen J, Mao H, Zeng X (2011) Twitter mood predicts the stock market. J Comput Sci 2: 1-8.
29. Golder SA, Macy MW (2011) Diurnal and seasonal mood vary with work, sleep, and daylength across diverse cultures. Science 333: 1878-1881.
30. Livne A, Simmons M, Adar E, Adamic L (2011) The party is over here: structure and content in the 2010 election. ICWSM 11: 17-21.
31. Ciulla F, Mocanu D, Baronchelli A, Goncalves B, Perra N, et al. (2012) Beating the news using social media: the case study of American Idol. EPJ Data Science 1.
32. Conover M, Goncalves B, Flammini A, Menczer F (2012) Partisan asymmetries in online political activity. EPJ Data Science 1.
33. Hannak A, Anderson E, Barrett LF, Lehmann S, Mislove A, et al. Tweetin' in the rain: Exploring societal-scale effects of weather on mood; 2012 June 2012; Dublin, Ireland.
34. Metaxas PT, Mustafaraj E (2012) Social Media and the Elections. Science 338: 472-473.
35. Yu S, Kak S (2012) A Survey of Prediction Using Social Media. arXiv preprint arXiv:12031647.
36. Mitchell L, Harris KD, Frank MR, Dodds PS, Danforth CM (2013) The Geography of Happiness: Connecting Twitter sentiment and expression, demographics, and objective characteristics of place. arXiv preprint arXiv:13023299.
37. Hogan J, Holland B (2003) Using theory to evaluate personality and job-performance relations: A socioanalytic perspective. Journal of Applied Psychology 88: 100–112.
38. Judge TA, Ilies R, Colbert AE (2004) Intelligence and leadership: A quantitative review and test of theoretical propositions. Journal of Applied Psychology 89.
39. House RJ, Aditya RN (1997) The social scientific study of leadership: Quo vadis? . Journal of Management 23: 409–473.
40. Ames DR, Flynn FJ (2007) What breaks a leader: The curvilinear relation between assertiveness and leadership. Journal of Personality and Social Psychology 92: 307–324.
41. Bass BM, Bass R, Bass BM (2008) The Bass handbook of leadership : theory, research, and managerial applications. New York: Free Press. xix, 1516 p. p.
42. King AJ, Johnson DD, Van Vugt M (2009) The origins and evolution of leadership. Curr Biol 19: R911-916.

43. Lin Y-R, Margolin D, Keegan B, Lazer D. Voices of victory: A computational focus group framework for tracking opinion shift in real time; 2013. International World Wide Web Conferences Steering Committee. pp. 737-748.
44. Weng L, Flammini A, Vespignani A, Menczer F (2012) Competition among memes in a world with limited attention. Scientific Reports 2.
45. Boersma P (2001) Praat, a system for doing phonetics by computer. Glottology International 5: 341-345.
46. Sasahara K, Hirata Y, Toyoda M, Kitsuregawa M, Aihara K (2013) Quantifying collective attention from tweet stream. PLoS One 8: e61823.
47. Pennacchiotti M, Popescu A-M. Democrats, republicans and starbucks afficionados: user classification in twitter; 2011. ACM. pp. 430-438.
48. Boutet A, Kim H, Yoneki E. What's in Your Tweets? I Know Who You Supported in the UK 2010 General Election; 2012.
49. Slatcher RB, Chung CK, Pennebaker JW, Stone LD (2007) Winning words: Individual differences in linguistic style among US presidential and vice presidential candidates. Journal of Research in Personality 41: 63-75.
50. Angus D, Smith A, Wiles J (2012) Human communication as coupled time series: Quantifying multi-participant recurrence. IEEE Transactions on Audio, Speech and Language Processing 20: 1795-1807.
51. Wu F, Huberman BA (2007) Novelty and collective attention. Proc Natl Acad Sci U S A 104: 17599-17601.
52. Crane R, Sornette D (2008) Robust dynamic classes revealed by measuring the response function of a social system. Proc Natl Acad Sci U S A 105: 15649-15653.
53. Lehmann J, Gonçalves B, Ramasco JJ, Cattuto C. Dynamical classes of collective attention in twitter; 2012. ACM. pp. 251-260.
54. Shockley K, Richardson DC, Dale R (2009) Conversation and coordinative structures. Topics in Cognitive Science 1: 305-319.
55. Dale R, Fusaroli R, Duran N, Richardson DC (2013) The self-organization of human interaction. Psychology of Learning and Motivation 59: 43-95.
56. Fusaroli R, Konvalinka I, Wallot S (2014) Analyzing social interactions: Promises and challenges of cross recurrence quantification analysis. Springer Proceedings in Mathematics & Statistics 103: 137-155.
57. Fusaroli R, Raczaszek-Leonardi J, Tylén K (2014) Dialog as interpersonal synergy. New Ideas in Psychology 32: 147-157.
58. Fusaroli R, Tylén K (in press) Investigating conversational dynamics: Interactive alignment, Interpersonal synergy, and collective task performance. Cognitive Science.
59. Strogatz SH (2003) Sync : the emerging science of spontaneous order. New York: Hyperion. viii, 338 p. p.
60. Kauffman S (1996) At Home in the Universe: The Search for the Laws of Self-Organization and Complexity. Oxford: Oxford University Press.
61. MacWhinney B (1999) The emergence of language. Mahwah, NJ: Taylor & Francis.
62. Loreto V, Steels L (2007) Social dynamics – Emergence of language. Nature Physics 3: 758–760.
63. Galantucci B, Garrod S (2010) Experimental semiotics: A new approach for studying the emergence and the evolution of human communication. Interaction Studies 11: 1-13.
18 of 21

64. Tylén K, Fusaroli R, Bundgaard P, Østergaard S (2013) Making sense together: A dynamical account of linguistic meaning making. Semiotica 194: 39-62.

# Supporting Information Legends

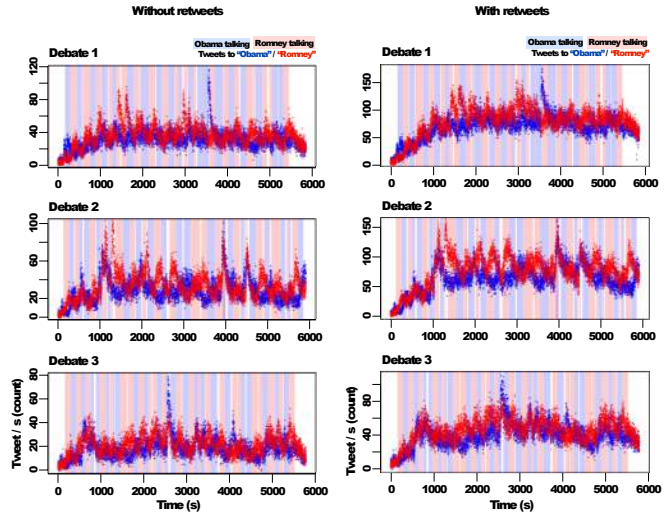

**Figure S1. Tweet rate and turn-taking the presidential debate: tweets vs. tweets+retweets** A comparison of tweet vs. total (tweet and retweet) rate (per second) of mentions to "Obama" and "Romney" across the debate. Patterns are highly similar, and retweets appear to happen very promptly following the volume of initial tweets.

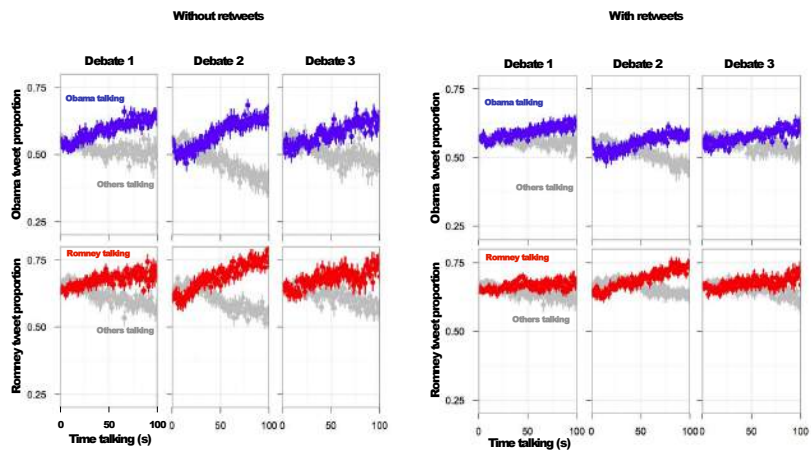

**Figure S2. Effects of taking and holding the ground on Twitter mentions: tweets vs. tweets+retweets** At the onset of speaking, results show that both the volume of tweets (on the left) and the total volume (tweets plus retweets, on the right) increase when that spoken turn is in the form of an interruption; this effect appears to be stronger in the original tweets.

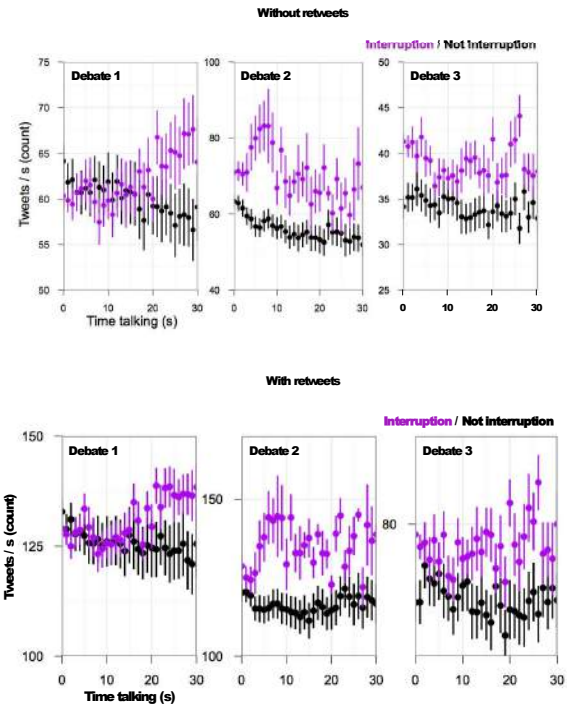

**Figure S3. Effects of interruptions on Twitter mentions: tweets vs. tweets+retweets** Original tweets (above) and total data (tweets plus retweets, below) both exhibit the interruption effect: during interruption by individuals during the debate, the raw tweet rate (per second) increases.

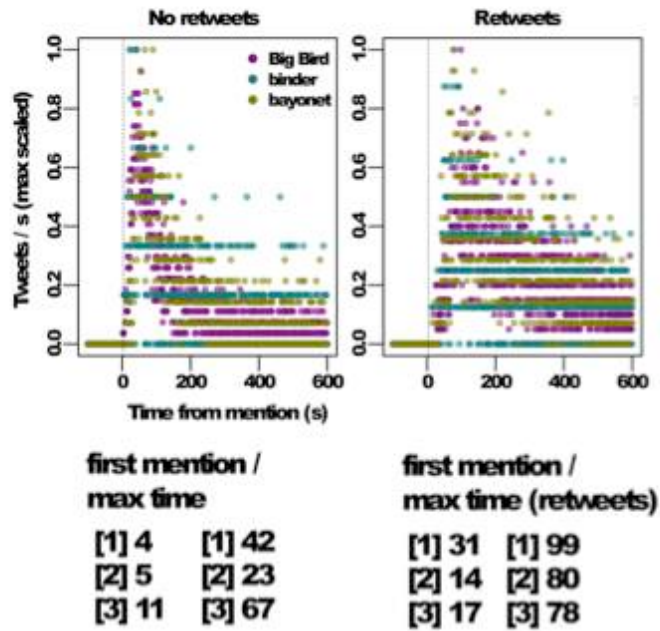

**Figure S4. The temporal profile of public attention to salient events: tweets vs. tweets+retweets** The left panel shows the original tweet data as displayed in the main paper. The retweets show a distinct time

delay that is still nevertheless highly similar in structure across all three pointed moments during the debates.

**File S1. Commented R code employed to run the analyses in the paper.** The file is a pdf containing the code used to run the analyses, commented for understandability, and the output of the code.